\documentclass[aps,pra,preprint,superscriptaddress,longbibliography]{revtex4-2}

\usepackage{amsmath}
\usepackage{amsthm}
\usepackage{amsfonts}
\usepackage{amssymb}
\usepackage{graphicx}
\usepackage{tikz}
\usepackage[pdftex]{hyperref}
\usepackage{longtable}
\usepackage{array}
\usepackage{dsfont}
\usepackage{bm}

\usepackage{etoolbox}

\begin{document}

    \title{Spectral contraction from Gaussian beams to Helmholtz waves}

    \author{A. {Dom\'inguez-Cruz}}
    \email[Contact author: ]{agustin.dominguez.cruz@gmail.com}
    \affiliation{Tecnol\'{o}gico de Monterrey, Monterrey 64849, M\'{e}xico.}

    \author{B.~M. Rodr\'iguez-Lara}
    \email[Contact author: ]{blas.rodriguez@gmail.com}
    \affiliation{Universidad Polit\'ecnica Metropolitana de Hidalgo, Tolcayuca, Hidalgo 43860, M\'{e}xico.}

    \date{\today}

    \begin{abstract}
    We introduce the ratio between the transverse Helmholtz length and the paraxial Gaussian-beam width as the physical parameter governing the contraction from separable Gaussian beams to transverse Helmholtz waves.  
    For a fixed transverse Helmholtz shell within the paraxial angular spectrum, the large-Rayleigh-range limit drives our optical scale ratio towards zero, while the higher-modal-order limit preserves the shell, transforming the isotropic and astigmatic paraxial oscillator eigenvalue problems into the transverse Helmholtz unit-shell eigenvalue problem.
    As our optical scale ratio vanishes, it realizes the algebraic In\"{o}n\"{u}-Wigner contraction of the isotropic Gaussian-beam $\mathrm{su}(2)$ algebra to the quadratic fixed-shell realization of the transverse Helmholtz $\mathrm{e}(2)$ algebra; conversely, finite nonzero values parametrize the Cayley-Klein deformation.
    Under the joint limit of vanishing scale ratio and higher modal order, paraxial Hermite-, Laguerre-, Ince-, and Boyer-Wolf-Gaussian beams undergo a spectral contraction in which their two modal numbers fix the shell and determine the single separation parameter labeling the corresponding Cartesian, Bessel, Mathieu, and Weber transverse Helmholtz waves, while preserving parity.
    Our spectral contraction extends to the wave operators, symmetry algebras, separation operators, separated equations, and separation spectra.
    \end{abstract}
    
    \maketitle
    \newpage 
    

\section{Introduction}
\label{sec:Sec1}
Structured light~\cite{Forbes2021p253} uses transverse spatial structure for modal encoding~\cite{Willner2015p66}, field shaping~\cite{Piccardo2022p013001}, and propagation engineering~\cite{MelladoVillasenor2026p023514}.
In free-space scalar optics~\cite{Levy2016p237}, the transverse Helmholtz equation supports propagation-invariant waves in fixed transverse-wavenumber shells, while the paraxial wave equation supports self-similar Gaussian beams in fixed total modal number oscillator subspaces.
The paraxial approximation provides their physical connection by writing a monochromatic field as a fast carrier multiplying a slow envelope and restricting its angular spectrum to wave vectors near the propagation axis~\cite{Born2013,Lax1975p1365,Agrawal1979p575}.
Plane~\cite{Born2013}, Bessel~\cite{Durnin1987p651,Jauregui2005p033411}, Mathieu~\cite{GutierrezVega2000p1493,RodriguezLara2008p033813}, and Weber~\cite{Bandres2004p44,RodriguezLara2009p055806} waves solve the Helmholtz equation in Cartesian, polar, elliptic, and parabolic coordinates.
Hermite-Gaussian~\cite{Born2013}, Laguerre-Gaussian~\cite{Allen1992p8185}, and Ince-Gaussian~\cite{Bandres2004p144} beams solve the paraxial equation through the isotropic oscillator~\cite{MoralesRodriguez2024p033523}, while Boyer-Wolf-Gaussian beams arise from the $2:1$ astigmatic oscillator~\cite{Boyer1975p2215} and have been introduced and observed recently in optics~\cite{Tschernig2024p5301}.
The group-theoretic separation program organizes these families through the Euclidean algebra $\mathrm{e}(2)$ for transverse Helmholtz waves and the angular momentum algebra $\mathrm{su}(2)$ for isotropic Gaussian beams~\cite{Miller2013}.

\begin{figure}[t]
    \centering
    \includegraphics[width=0.75\textwidth]{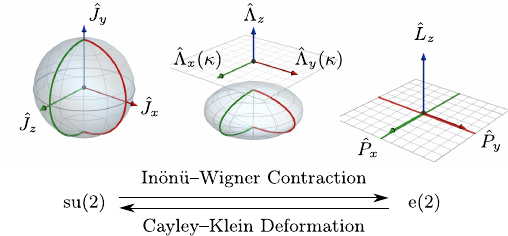}
    \caption{
    Algebraic passage for the Cartesian and polar families.
    The In\"on\"u-Wigner contraction flattens the $\mathrm{su}(2)$ sphere into the $\mathrm{e}(2)$ plane; the Cayley-Klein deformation reverses it.
    }
    \label{fig:Fig1}
\end{figure}

Beyond the physical connection provided by the paraxial approximation, several approaches bridge Gaussian and Helmholtz solutions at other levels.
Nonparaxial expansions give higher-order corrections~\cite{Lax1975p1365,Agrawal1979p575,Cerjan2011p2253}, family-specific comparisons and asymptotics connect selected Gaussian beams with limiting wave families~\cite{MendozaHernandez2015p3739,Bhargava2024p5320,Tschernig2024p5301}, and complex-source models regularize Bessel-Gauss propagation~\cite{April2011p2100,GutierrezVega2005p289,NguyenThi2025p15025}.
The algebraic In\"on\"u-Wigner contraction gives the broadest passage by contracting the $\mathrm{su}(2)$ Lie algebra to $\mathrm{e}(2)$, flattening the finite Gaussian-mode sphere into the Helmholtz plane (Fig.~\ref{fig:Fig1})~\cite{Inonu1953p510,Wolf2018p976,Nieto1998p3875} and connecting the Cartesian and polar families.
A physical framework connecting all four separable Gaussian-beam families with their transverse Helmholtz-wave counterparts at the coordinate, separation operator, separated differential equation, and separation spectrum levels remains missing.

Here, we construct such a framework by comparing the transverse Helmholtz length with the Gaussian-beam transverse scale.
We formulate transverse Helmholtz waves, Sec.~\ref{sec:Sec2}, and paraxial Gaussian beams, Sec.~\ref{sec:Sec3}, on their respective dimensionless transverse scales and organize the four separable families through their parity labels, modal numbers, and separation operators.
In Sec.~\ref{sec:Sec4}, we define our optical scale ratio as the ratio between these transverse scales and show that, for a fixed transverse Helmholtz shell chosen within the paraxial angular spectrum, the large-Rayleigh-range limit increases the Gaussian width relative to the transverse Helmholtz length and drives our optical scale ratio towards zero.
We couple this optical limit to the higher-modal-order limit so that the scaled paraxial oscillator eigenvalue remains on the fixed shell, producing a spectral contraction in which the two modal numbers of each Gaussian beam fix the shell and determine the single separation parameter of its corresponding transverse Helmholtz wave.
We establish our contraction through the wave operators, separation operators, quadratic fixed-shell realization of $\mathrm{e}(2)$, separated differential equations, and separation spectra of the Cartesian, polar, elliptic, and parabolic families; Sec.~\ref{sec:Sec5} summarizes our contraction's physical and algebraic consequences.

\section{Helmholtz waves}
\label{sec:Sec2}
We use a complex scalar monochromatic field propagating in the $z$ direction,
\begin{align}
    \mathcal{E}_{\lambda_{\nu}}^{(\nu)}(x,y,z,t) = \Phi_{\lambda_{\nu}}^{(\nu)}(x,y)
      e^{i(k_{z} z-\omega t)},
\end{align}
with the axial wavenumber $k_{z}$, the angular frequency $\omega$, and the real propagation-invariant transverse spatial mode $\Phi_{\lambda_{\nu}}^{(\nu)}(x,y)$.
The transverse wavenumber $k_{\perp}^{2} = k^{2} - k_{z}^{2}$, with the medium wavenumber $k = \omega / v$ and the light speed $v$ in the medium, sets a transverse scale, which we use to define the dimensionless coordinates,
\begin{align}
    Q_{a} = k_{\perp} a, \qquad
    a\in\{x,y\},
\end{align}
and the dimensionless position and linear-momentum operators acting on the transverse Helmholtz wave mode,
\begin{align}
    \begin{aligned}
        \hat{Q}_{a} \Phi_{\lambda_{\nu}}^{(\nu)} =&~ Q_{a} \Phi_{\lambda_{\nu}}^{(\nu)}, \\
        \hat{P}_{a} \Phi_{\lambda_{\nu}}^{(\nu)} =&~ - i \partial_{Q_{a}} \Phi_{\lambda_{\nu}}^{(\nu)},
    \end{aligned}
\end{align}
understood as multiplication and differential operators acting on the transverse modal space.
Under our scaling, the transverse Helmholtz equation takes the unit-shell form,
\begin{align}
    \hat{P}^{2} \Phi_{\lambda_{\nu}}^{(\nu)} &= \Phi_{\lambda_{\nu}}^{(\nu)}, \qquad
    \hat{P}^{2} = \hat{P}_{x}^{2} + \hat{P}_{y}^{2},
\end{align}
where $\nu \in \{\mathrm{C}, \mathrm{B}, \mathrm{M}, \mathrm{W}\}$ labels Cartesian, Bessel, Mathieu, and Weber waves, while $\lambda_{\nu}$ is the set of mode indices identifying each real separable Helmholtz wave within its family.

Each real separable transverse Helmholtz wave simultaneously diagonalizes the squared transverse momentum $\hat{P}^{2}$ and one parity-preserving separation operator.
Real parity-adapted Cartesian plane-wave combinations carry the Cartesian parity vector $\boldsymbol{\sigma} = (\sigma_x,\sigma_y)$, with $\sigma_a \in \{\mathrm{e},\mathrm{o}\}$, while Bessel, Mathieu, and Weber waves share the parity label $\sigma \in \{\mathrm{e},\mathrm{o}\}$.
Their additional separation labels are given by the Cartesian separation angle $\vartheta \in [0,\pi/2]$, which provides the squared transverse-momentum components $\cos^{2}\vartheta$ and $\sin^{2}\vartheta$, and the Cartesian separation operator 
\begin{align}
    \hat{A}_{\mathrm{C}} \Phi^{(\mathrm{C})}_{\boldsymbol{\sigma},\vartheta} = \frac{1}{4}\cos(2\vartheta)\Phi^{(\mathrm{C})}_{\boldsymbol{\sigma},\vartheta}, \qquad
    \hat{A}_{\mathrm{C}} = \frac{1}{4} \left( \hat{P}_{x}^{2} - \hat{P}_{y}^{2} \right),
\end{align}
for real parity-adapted Cartesian waves, $\lambda_{\mathrm{C}}=(\boldsymbol{\sigma},\vartheta)$, with all four Cartesian parity vectors allowed for $\vartheta\in(0,\pi/2)$, while only $\sigma_{y}=\mathrm{e}$ gives nonzero modes at $\vartheta=0$ and only $\sigma_{x}=\mathrm{e}$ gives nonzero modes at $\vartheta=\pi/2$; the azimuthal order $\ell$ and the squared axial orbital angular momentum
\begin{align}
    \hat{L}_{z}^{2} \Phi^{(\mathrm{B})}_{\sigma,\ell} = \ell^{2}  \Phi^{(\mathrm{B})}_{\sigma,\ell}, \qquad
    \hat{L}_{z} = \hat{Q}_{x} \hat{P}_{y} - \hat{Q}_{y} \hat{P}_{x},
\end{align}
for Bessel waves, $\lambda_{\mathrm{B}}=(\sigma,\ell)$, with $\ell \in \mathbb{N}_{0}=\{0,1,2,\dots\}$ for even modes and $\ell \in \mathbb{N}=\{1,2,\dots\}$ for odd modes; the elliptic order $m$ and the Mathieu separation operator 
\begin{align}
    \hat{A}_{\mathrm{M}} \Phi^{(\mathrm{M})}_{\sigma,m} = a_{\sigma,m}\left( \frac{1}{4} Q_{f}^2 \right) \Phi^{(\mathrm{M})}_{\sigma,m}, \qquad
    \hat{A}_{\mathrm{M}} = \hat{L}_{z}^{2} + \frac{1}{2} Q_{f}^{2} \left(\hat{P}_{x}^{2}-\hat{P}_{y}^{2}\right),
\end{align}
for Mathieu waves, $\lambda_{\mathrm{M}}=(\sigma,m)$, with $m \in \mathbb{N}_{0}$ for even waves and $m \in \mathbb{N}$ for odd waves; and the continuous parameter $b \in \mathbb{R}$ and the Weber separation operator
\begin{align}
    \hat{B}_{\mathrm{W}} \Phi^{(\mathrm{W})}_{\sigma,b} = b \Phi^{(\mathrm{W})}_{\sigma,b}, \qquad
    \hat{B}_{\mathrm{W}} = \frac{1}{2} \left\{ \hat{L}_{z}, \hat{P}_{y} \right\},
\end{align}
for Weber waves, $\lambda_{\mathrm{W}}=(\sigma,b)$, which sets their parabolic order.
Here $Q_{f}=k_{\perp}f$ is the scaled elliptic half-interfocal distance, $a_{\sigma, m}(Q_f^2/4)$ is the Mathieu separation constant~\cite{NISTBOOK}, and $\{\hat{A}, \hat{B} \} = \hat{A} \hat{B} + \hat{B} \hat{A}$ is the anticommutator.
The ideal transverse Helmholtz waves are generalized eigenfunctions with Dirac-$\delta$ normalization in their continuous spectral labels and Kronecker normalization in their discrete labels.

\section{Paraxial beams}
\label{sec:Sec3}
We use a complex scalar monochromatic field propagating in the $z$ direction,
\begin{align}
    \mathcal{E}_{\lambda_{\mu}}^{(\mu)}(x,y,z,t)=&~ \frac{\sqrt{2}}{w(z)} \Psi_{\lambda_{\mu}}^{(\mu)}(q_{x},q_{y})
    e^{i\left[kz-\omega t+\frac{k}{2R(z)}\left(x^{2}+y^{2}\right)-(N+1)\zeta(z)\right]},
\end{align}
with the Gaussian-beam width $w(z) = w_{0} \sqrt{ 1 + (z / z_{R})^{2}}$, the waist $w_{0}$, the Rayleigh range $z_{R} = k w_{0}^{2} / 2$, the wavefront radius $R(z) = z [ 1 + (z_{R} / z)^{2}]$, the Gouy phase $\zeta(z) = \tan^{-1}( z / z_{R} )$, the modal order $N$, and the real self-similar transverse spatial mode $\Psi_{\lambda_{\mu}}^{(\mu)}(q_{x},q_{y})$, containing a real Gaussian envelope and a family-dependent modal factor.
The Gaussian-beam width sets a transverse scale, which we use to define the dimensionless coordinates,
\begin{align}
    q_{a} = \frac{\sqrt{2}}{w(z)} a, \qquad
    a \in \{ x, y\},
\end{align}
and the dimensionless position and linear-momentum operators acting on the transverse Gaussian beam mode,
\begin{align}
    \begin{aligned}
        \hat{q}_{a}\Psi_{\lambda_{\mu}}^{(\mu)} =&~ q_{a}\Psi_{\lambda_{\mu}}^{(\mu)}, \\
        \hat{p}_{a}\Psi_{\lambda_{\mu}}^{(\mu)} =&~ -i\partial_{q_{a}}\Psi_{\lambda_{\mu}}^{(\mu)},
    \end{aligned}
\end{align}
understood as multiplication and differential operators acting on the transverse modal space.
Under our scaling, the paraxial wave equation takes the isotropic oscillator eigenvalue form for $\mu\in\{\mathrm{HG},\mathrm{LG},\mathrm{IG}\}$,
\begin{align}
        \hat{N}\Psi_{\lambda_{\mu}}^{(\mu)} = N \Psi_{\lambda_{\mu}}^{(\mu)}, \qquad
        \hat{N} = \hat{N}_{x}+\hat{N}_{y}, 
\end{align}
where $\mu$ labels the Gaussian-beam family, while $\lambda_{\mu}$ is the set of mode indices identifying each real separable Gaussian beam within its family, and the Cartesian modal number operators
\begin{align}
        \hat{N}_{a} =&~ \frac{1}{2}\left(\hat{p}_{a}^{2}+\hat{q}_{a}^{2}\right)-\frac{1}{2}.
\end{align}

Each real separable paraxial Gaussian beam simultaneously diagonalizes the modal-number operator and one parity-preserving separation operator.
While each Helmholtz wave on the fixed unit shell carries parity and one separation label, each Gaussian beam carries parity and two modal numbers.
Hermite-Gaussian beams carry the Cartesian parity vector $\boldsymbol{\sigma} = (\sigma_x,\sigma_y)$, with $\sigma_a \in \{\mathrm{e},\mathrm{o}\}$, while Laguerre-, Ince-, and Boyer-Wolf-Gaussian beams share the parity label $\sigma \in \{\mathrm{e},\mathrm{o}\}$.
Their two modal numbers and separation operators are given by the Cartesian modal numbers $(n_x, n_y)$ and the oscillator-difference separation operator
\begin{align}
	\hat{J}_{z} \Psi_{\boldsymbol{\sigma},n_x,n_y}^{(\mathrm{HG})} = \frac{n_{x} - n_{y}}{2} \Psi_{\boldsymbol{\sigma},n_x,n_y}^{(\mathrm{HG})}, \qquad
	\hat{J}_{z} = \frac{1}{2} \left( \hat{N}_{x} - \hat{N}_{y} \right),
\end{align}
for Hermite-Gaussian beams, $\lambda_{\mathrm{HG}} = (\boldsymbol{\sigma}, n_{x}, n_{y})$, with the parity vector determined by the parity of the modal numbers $n_{a} \in \mathbb{N}_{0}$; the radial and azimuthal modal numbers $(p, \ell)$ and the squared angular-momentum separation operator
\begin{align}
	4 \hat{J}_{y}^{2}\Psi_{\sigma, p, \ell}^{(\mathrm{LG})} = \ell^{2} \Psi_{\sigma,p,\ell}^{(\mathrm{LG})}, \qquad
	\hat{J}_{y} = \frac{1}{2} \left( \hat{q}_{x}\hat{p}_{y} - \hat{q}_{y}\hat{p}_{x} \right),
\end{align}
for Laguerre-Gaussian beams, $\lambda_{\mathrm{LG}}=(\sigma, p, \ell )$, with $p \in\mathbb N_{0}$ and $\ell \in\mathbb N_{0}$ for even modes and $\ell \in\mathbb N$ for odd modes; the total and elliptic modal numbers $(N, m)$ and the Ince separation operator
\begin{align}
	\hat{\mathcal{J}}_{\mathrm{IG}} \Psi_{\sigma,N,m}^{(\mathrm{IG})} = \alpha_{\sigma, N, m}(q_{f}^{2}) \Psi_{\sigma, N, m}^{(\mathrm{IG})}, \qquad
	\hat{\mathcal{J}}_{\mathrm{IG}} = 4 \hat{J}_{y}^{2} + 2 q_{f}^{2} \hat{J}_{z},
\end{align}
for Ince-Gaussian beams, $\lambda_{\mathrm{IG}} = ( \sigma, N, m)$, with $N \in \mathbb{N}_{0}$, $0 \leq m \leq N$ for even modes, and $1 \leq m \leq N$ for odd modes, subject to $N - m \in 2 \mathbb{N}_{0}$.
Here $q_{f} = \sqrt{2} f / w_{0}$, with waist half-interfocal distance $f$, is the dimensionless elliptic half-interfocal distance, and $\alpha_{\sigma,N,m}(q_{f}^{2})$ is the Ince separation constant.
The isotropic modal order
\begin{align}
	N = n_{x} + n_{y} = 2 p + \ell = n_{+} + n_{-}
\end{align}
fixes a finite $(N+1)$-dimensional spin-$j=N/2$ representation of $\mathrm{su}(2)$~\cite{MoralesRodriguez2024p1489,TorresLeal2024p063716}.
Diagonal Hermite-Gaussian beams with modal numbers $(n_{+},n_{-})$ and diagonal parity vector $\boldsymbol{\sigma}_{\pm}=(\sigma_{+},\sigma_{-})$ complete the generator triple,
\begin{align}
	\hat{J}_{x} \Psi_{\boldsymbol{\sigma}_{\pm},n_{+},n_{-}}^{(\mathrm{DHG})} = \frac{n_{+} - n_{-}}{2} \Psi_{\boldsymbol{\sigma}_{\pm},n_{+},n_{-}}^{(\mathrm{DHG})}, \qquad
	\hat{J}_{x} =\frac{1}{2}\left( \hat{p}_{x} \hat{p}_{y} + \hat{q}_{x} \hat{q}_{y} \right),
\end{align}
with $n_{\pm} \in \mathbb{N}_{0}$, the parity vector $\boldsymbol{\sigma}_{\pm}$ determined by the parity of $n_{\pm}$, $[ \hat{N}, \hat{J}_{a} ] = 0$, and $[\hat{J}_{a}, \hat{J}_{b} ] = i \varepsilon_{abc} \hat{J}_{c}$ for $a, b, c \in \{x, y, z\}$, with $\varepsilon_{abc}$ being the Levi-Civita symbol.

Boyer-Wolf-Gaussian beams are transverse eigenmodes of a $2:1$ anisotropic gradient-index (GRIN) medium~\cite{Tschernig2024p5301} and do not propagate self-similarly in free space.
Thus, we work with their complex scalar monochromatic field at the waist reference plane,
\begin{align}
    \mathcal{E}_{\sigma,n,r}^{(\mathrm{BWG})}(x,y,0,t) = \frac{\sqrt{2}}{w_{0}} \Psi_{\sigma,n,r}^{(\mathrm{BWG})}(q_x,q_y)e^{-i\omega t},
\end{align}
whose real transverse spatial mode $\Psi_{\sigma,n,r}^{(\mathrm{BWG})}(q_x,q_y)$ solves the $2:1$ astigmatic oscillator eigenvalue problem,
\begin{align}
    \hat{N}^{(2:1)}\Psi_{\sigma,n,r}^{(\mathrm{BWG})} = n\Psi_{\sigma,n,r}^{(\mathrm{BWG})}, \qquad
    \hat{N}^{(2:1)} = \hat{N}_{x}^{(2)}+\hat{N}_{y},
\end{align}
with the frequency-doubled Cartesian modal-number operator
\begin{align}
    \hat{N}_{x}^{(2)} = \frac{1}{2}\left(\hat{p}_{x}^{2}+4\hat{q}_{x}^{2}\right)-1.
\end{align}
Each real separable mode simultaneously diagonalizes $\hat{N}^{(2:1)}$ and the parity-preserving separation operator~\cite{Boyer1975p2215},
\begin{align}
    \hat{\mathcal{J}}_{\mathrm{BWG}}^{(2:1)} \Psi_{\sigma,n,r}^{(\mathrm{BWG})} = \beta_{n,r} \Psi_{\sigma,n,r}^{(\mathrm{BWG})}, \qquad
    \hat{\mathcal{J}}_{\mathrm{BWG}}^{(2:1)} = \hat{q}_{x} \hat{q}_{y}^{2} -\left\{ \hat{J}_{y}, \hat{p}_{y} \right\},
\end{align}
where $\beta_{n,r}$ is the separation constant, $n \in \mathbb{N}_0$ is the modal order, and $r \in \mathcal R_{n} = \{ -\frac{1}{2} \lfloor n/2 \rfloor, -\frac{1}{2} \lfloor n/2 \rfloor + 1, \ldots, \frac{1}{2} \lfloor n/2 \rfloor\}$ is the branch index, with $\lfloor \cdot \rfloor$ being the floor function. 
The label set is $\lambda_{\mathrm{BWG}}=(\sigma,n,r)$, where the parity of the modal number $n$ determines the parity $\sigma$ of the beam.

\section{Spectral contraction}
\label{sec:Sec4}
At fixed $z$, we connect the Helmholtz and paraxial descriptions,
\begin{align}
    \begin{aligned}
        \hat{q}_{a} =&~ \sqrt{\kappa} ~ \hat{Q}_{a}, \\
        \hat{p}_{a} =&~ \frac{1}{\sqrt{\kappa}} ~ \hat{P}_{a},
    \end{aligned}
\end{align}
through our optical scale ratio,
\begin{align}
    \begin{aligned}
        \kappa(z) =&~ \frac{2}{k_{\perp}^{2} w^{2}(z)}, \\
        =&~ \left( \frac{k}{k_{\perp}} \right)^{2} \frac{1}{k z_{R}} \left[ 1 + \left( \frac{z}{z_{R}} \right)^{2} \right]^{-1}.
    \end{aligned}
\end{align}
Its square root, $\sqrt{\kappa} = \sqrt{2} / [ k_{\perp} w(z)]$, is the ratio between the transverse Helmholtz length $k_{\perp}^{-1}$ and the Gaussian transverse scale $w(z)/\sqrt{2}$.
To connect a paraxial Gaussian beam with a transverse Helmholtz wave, we choose a fixed transverse wavenumber satisfying the near-axis condition $k_{\perp}/k \ll 1$, which places the corresponding transverse Helmholtz shell within the paraxial angular spectrum.
We take the large-Rayleigh-range limit $z_{R}\to\infty$, which increases the Gaussian width relative to the transverse Helmholtz length, makes the Gaussian envelope flat, and suppresses its transverse rescaling and wavefront curvature at every propagation plane $z$.
This limit recovers the propagation-invariant transverse geometry and drives our optical scale ratio towards zero, $\kappa \to 0^{+}$.
For Boyer-Wolf-Gaussian beams, whose free-space propagation is not self-similar, we apply the transverse scaling at the waist reference plane.

In our scaled variables $\hat{q}_a= \kappa^{1/2} \hat{Q}_a$ and $\hat{p}_a=\kappa^{-1/2}\hat{P}_a$, the isotropic and $2:1$ astigmatic modal-number operators,
\begin{align}
    \begin{aligned}
    \hat{N} & =  \frac{1}{2\kappa} \left[ \hat{P}^2 + \kappa^2 \left( \hat{Q}_{x}^{2} + \hat{Q}_{y}^{2} \right) - 2 \kappa \right], \\
    \hat{N}^{(2:1)} & = \frac{1}{2 \kappa} \left[ \hat{P}^{2} + \kappa^{2} \left( 4 \hat{Q}_{x}^{2} + \hat{Q}_{y}^{2} \right) - 3 \kappa \right],
    \end{aligned}
\end{align}
substituted into the eigenvalue equations $\hat{N}\Psi_{\lambda_{\mu}}^{(\mu)} = N \Psi_{\lambda_{\mu}}^{(\mu)}$ and  $\hat{N}^{(2:1)}\Psi_{\lambda_{\mathrm{BWG}}}^{(\mathrm{BWG})} = n\Psi_{\lambda_{\mathrm{BWG}}}^{(\mathrm{BWG})}$ provide, after multiplication by $2 \kappa$, the scaled eigenvalue equations,
\begin{align}\label{eq:QHOs}
    \begin{aligned}
        \left[ \hat{P}^{2} + \kappa^{2} ( \hat{Q}_{x}^{2} + \hat{Q}_{y}^{2} ) - 2\kappa
        \right] \Psi_{\lambda_{\mu}}^{(\mu)} =&~ 2 \kappa N \Psi_{\lambda_{\mu}}^{(\mu)},
        \\
        \left[ \hat{P}^{2} + \kappa^{2}( 4 \hat{Q}_{x}^{2} + \hat{Q}_{y}^{2} ) - 3\kappa \right] \Psi_{\lambda_{\mathrm{BWG}}}^{(\mathrm{BWG})} =&~ 2\kappa n \Psi_{\lambda_{\mathrm{BWG}}}^{(\mathrm{BWG})}.
    \end{aligned}
\end{align}
At the fixed transverse coordinate $Q_{a}$, the oscillator potentials vanish quadratically with $\kappa$, while the constant terms vanish linearly, leaving the squared transverse-momentum operator $\hat{P}^{2}$.
We preserve the unit-shell eigenvalue of the transverse Helmholtz eigenvalue problem by coupling the vanishing optical scale ratio to a diverging modal order,
\begin{align}
    \begin{aligned}
        \lim_{\substack{\kappa \to 0^{+} \\ N \to \infty}} 2 \kappa N =&~ 1, \\ 
        \lim_{\substack{\kappa \to 0^{+} \\ n \to \infty}} 2 \kappa n  =&~ 1.
    \end{aligned}
\end{align}
Under these limits, the paraxial oscillator eigenvalue problems reduce to the transverse Helmholtz unit-shell equations,
\begin{align}
    \begin{aligned}
        \lim_{\substack{\kappa \to 0^{+} \\ N \to \infty}}  \left[ \hat{P}^{2} + \kappa^{2} ( \hat{Q}_{x}^{2} + \hat{Q}_{y}^{2} ) - 2 \kappa \left( N + 1 \right) \right] \Psi_{\lambda_{\mu}}^{(\mu)} =&~ \left[ \hat{P}^{2}  - 1 \right] \Psi_{\lambda_{\mu}}^{(\mu)}, \\
        \lim_{\substack{\kappa \to 0^{+} \\ n \to \infty}} \left[ \hat{P}^{2} + \kappa^{2}( 4 \hat{Q}_{x}^{2} + \hat{Q}_{y}^{2} ) - \kappa \left( 2 n + 3 \right) \right] \Psi_{\lambda_{\mathrm{BWG}}}^{(\mathrm{BWG})} =&~ \left[ \hat{P}^{2}  - 1 \right] \Psi_{\lambda_{\mathrm{BWG}}}^{(\mathrm{BWG})},
    \end{aligned}
\end{align}
connecting paraxial Gaussian beams with their corresponding transverse Helmholtz waves through a spectral contraction that couples increasing Gaussian-beam width to higher modal order, while their two modal numbers fix the shell and determine the single separation parameter labeling the corresponding wave.

\subsection{Cayley-Klein Deformation and In\"{o}n\"{u}-Wigner Contraction}

Our optical scale ratio $\kappa$ parametrizes the Cayley-Klein deformation~\cite{Herranz1994p2515} of the isotropic Gaussian-beam $\mathrm{su}(2)$ generators, 
in the sense of a one-parameter family of algebras whose $\kappa \to 0^{+}$ limit realizes the In\"{o}n\"{u}-Wigner contraction to the transverse Helmholtz $\mathrm{e}(2)$ algebra~\cite{Wolf2018p976,Nieto1998p3875},
\begin{align}
    \begin{aligned}
        \hat{\Lambda}_{x}(\kappa) =&~ \kappa \hat{J}_{z} = \frac{1}{4}( \hat{P}_{x}^{2} - \hat{P}_{y}^{2} ) + \frac{1}{4} \kappa^{2} \left( \hat{Q}_{x}^{2} - \hat{Q}_{y}^{2} \right), \\
        \hat{\Lambda}_{y}(\kappa) =&~ \kappa \hat{J}_{x} = \frac{1}{2} \hat{P}_{x} \hat{P}_{y} + \frac{1}{2} \kappa^{2} \hat{Q}_{x} \hat{Q}_{y}, \\
        \hat{\Lambda}_{z} =&~ \hat{J}_{y} = \frac{1}{2} \hat{L}_{z},
    \end{aligned}
\end{align}
which close the curvature-dependent algebras,
\begin{align}
    \begin{aligned}
        \left[ \hat{\Lambda}_{z}, \hat{\Lambda}_{x} \right] =&~ i \hat{\Lambda}_{y},\\
        \left[ \hat{\Lambda}_{z}, \hat{\Lambda}_{y} \right] =&~ - i \hat{\Lambda}_{x},\\
        \left[ \hat{\Lambda}_{x}, \hat{\Lambda}_{y} \right] =&~ i \kappa^{2} \hat{\Lambda}_{z}.
    \end{aligned}
\end{align}
At finite $\kappa$, this algebra is isomorphic to $\mathrm{su}(2)$, with the Casimir operator
\begin{align}
    \hat{\Lambda}^{2}(\kappa) = \hat{\Lambda}^{2}_{x}(\kappa) + \hat{\Lambda}^{2}_{y}(\kappa) + \kappa^{2}\hat{\Lambda}^{2}_{z} = \kappa^{2}\hat{\boldsymbol{J}}^{2},
\end{align}
and eigenvalue in the spin-$j=N/2$ representation~\cite{MoralesRodriguez2024p033523,TorresLeal2024p063716}
\begin{align}
    \hat{\Lambda}^{2}(\kappa)\Psi_{\lambda_{\mu}}^{(\mu)} = \frac{1}{4}\kappa^{2}N\left(N+2\right)\Psi_{\lambda_{\mu}}^{(\mu)},
\end{align}
while the limit $\kappa\to0^{+}$ gives the generators
\begin{align}
    \begin{aligned}
        \hat{\Lambda}_{x}(0) =&~ \frac{1}{4} \left( \hat{P}_{x}^{2}-\hat{P}_{y}^{2}\right) = \hat{A}_{\mathrm{C}}, \\
        \hat{\Lambda}_{y}(0) =&~ \frac{1}{2}\hat{P}_{x}\hat{P}_{y}, \\ 
        2\hat{\Lambda}_{z} =&~ \hat{L}_{z},
    \end{aligned}
\end{align}
which, under our spectral contraction, form a quadratic fixed-shell realization of $\mathrm{e}(2)$ as the finite-scale Casimir eigenvalue contracts to the Casimir eigenvalue of the Helmholtz unit shell,
\begin{align}
    \begin{aligned}
        \lim_{\substack{\kappa \to 0^{+} \\ N \to \infty}} \hat{\Lambda}^{2}(\kappa) = \frac{1}{16}\left(\hat{P}_{x}^{2}+\hat{P}_{y}^{2}\right)^{2} = \frac{1}{16}.
    \end{aligned}
\end{align}
Our spectral contraction realizes the In\"{o}n\"{u}-Wigner contraction~\cite{Inonu1953p510} through the physical scale ratio $\kappa$, giving the paraxial-Helmholtz connection at the algebraic and representation levels.

\subsection{Separation operators}
At the separation-operator level, the limit $\kappa \to 0^{+}$ sends $\hat{\Lambda}_{x}(\kappa)$ to the Cartesian separator $\hat{A}_{\mathrm{C}}$, diagonalized by the real parity-adapted Cartesian-wave combinations $\Phi^{(\mathrm{C})}_{\boldsymbol{\sigma},\vartheta}$ with the eigenvalue $\cos(2\vartheta)/4$, while $4\hat{\Lambda}_{z}^{2}=\hat{L}_{z}^{2}$ gives the parity-preserving polar separator diagonalized by the real Bessel waves $\Phi^{(\mathrm{B})}_{\sigma,\ell}$ with the eigenvalue $\ell^{2}$. 
The elliptic and parabolic Gaussian-beam separators take the scaled forms
\begin{align}
    \begin{aligned}
        \hat{\mathcal{J}}_{\mathrm{IG}}(\kappa) =&~ \hat{A}_{\mathrm{M}} + \frac{1}{2} \kappa^{2} Q_{f}^{2} \left( \hat{Q}_{x}^{2} - \hat{Q}_{y}^{2} \right), \\
        -\sqrt{\kappa} \hat{\mathcal{J}}_{\mathrm{BWG}}^{(2:1)}(\kappa) =&~ \hat{B}_{\mathrm{W}}
        - \kappa^{2} \hat{Q}_{x} \hat{Q}_{y}^{2},
    \end{aligned}
\end{align}
which contract to the Mathieu and Weber Helmholtz separators as $\kappa \to 0^{+}$.
For these modal contractions, we use the waist plane and write $\kappa = \kappa(0)$, so $q_{f}^{2} = \kappa Q_{f}^{2}$.
The corresponding elliptic and parabolic coordinates scale according to
\begin{align}
    \begin{aligned}
        q_{s} =&~ Q_{s}, \qquad
        & \hat{p}_{q_{s}} =&~ \hat{P}_{Q_{s}}, \qquad
        & s \in&~ \{\xi,\eta\}, \\
        q_{c} =&~ \kappa^{1/4}Q_{c}, \qquad
        & \hat{p}_{q_{c}} =&~ \kappa^{-1/4}\hat{P}_{Q_{c}}, \qquad
        & c \in&~ \{u,v\},
    \end{aligned}
\end{align}
where the elliptic definition $x + i y = f \cosh( \xi + i \eta)$ leaves the elliptic coordinates unchanged under the transverse scaling, while the parabolic definition $x + i y=(u + i v )^{2}/2$ produces the quarter-power scaling.
Although the Boyer-Wolf separation operator commutes with $\hat{N}^{(2:1)}$ and not with the isotropic modal-number operator $\hat{N}$, the two rescaled modal-number operators share the same transverse Helmholtz limit,
\begin{align}
    \lim_{\kappa \to 0^{+}} 2 \kappa \hat{N} = \lim_{\kappa \to 0^{+}} 2 \kappa \hat{N}^{(2:1)}
    = \hat{P}^{2},
\end{align}
while the corresponding nonvanishing commutator contracts to
\begin{align}
    \lim_{\kappa\to0^{+}}
    \left[
        2\kappa\hat{N},
        -\sqrt{\kappa}\hat{\mathcal{J}}_{\mathrm{BWG}}^{(2:1)}(\kappa)
    \right]
    = \left[\hat{P}^{2},\hat{B}_{\mathrm{W}}\right]
    = 0,
\end{align}
with the real Weber waves $\Phi_{\sigma,b}^{(\mathrm{W})}$ as their simultaneous eigenfunctions.
Our spectral contraction connects the Gaussian-beam separators with their Helmholtz-wave counterparts at the separation-operator level.

\subsection{Separated differential equations and spectra}
\begin{figure}[t]
    \centering
    \includegraphics[width=0.75\textwidth]{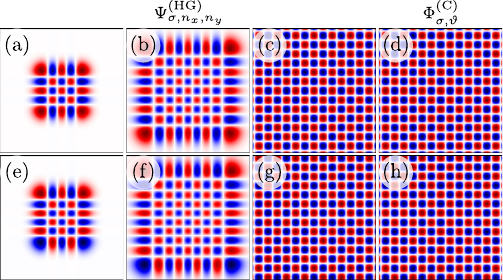}
    \caption{
    Real Hermite-Gauss to parity-adapted Cartesian-wave spectral contraction for $\vartheta=\pi/4$.
    The top row shows the even-even sector and the bottom row shows the even-odd sector.
    The first three columns follow the parity-compatible Cartesian modal-number sequences, and the final column shows the corresponding transverse Helmholtz wave.
    }
    \label{fig:Fig2-HG}
\end{figure}

\begin{figure}[t]
    \centering
    \includegraphics[width=0.75\textwidth]{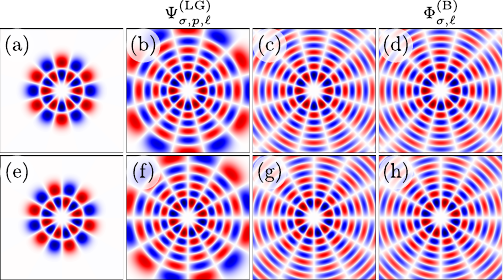}
    \caption{
    Real Laguerre-Gauss to Bessel spectral contraction for fixed azimuthal order $\ell=6$.
    The top and bottom rows show the even and odd sectors.
    The first three columns use $p=(N-6)/2$ with $N=8,16,104$, and the final column shows the corresponding Bessel wave.
    }
    \label{fig:Fig3-LG}
\end{figure}

\begin{figure}[t]
    \centering
    \includegraphics[width=0.75\textwidth]{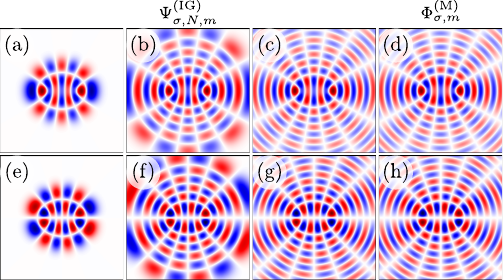}
    \caption{
    Real Ince-Gauss to Mathieu spectral contraction for fixed elliptic order $m=6$ and scaled half-interfocal distance $Q_{f}=8$.
    The top and bottom rows show the even and odd sectors.
    The first three columns use $N=8,16,104$, and the final column shows the corresponding Mathieu wave.
    }
    \label{fig:Fig4-IG}
\end{figure}

\begin{figure}[t]
    \centering
    \includegraphics[width=0.75\textwidth]{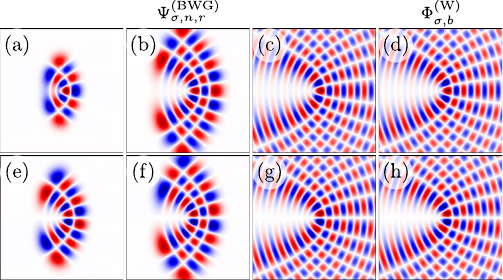}
    \caption{
    Real Boyer-Wolf-Gauss to Weber spectral contraction for fixed separation parameter $b=2$.
    The top and bottom rows show the even and odd sectors.
    The first three columns follow the parity-compatible modal-order and branch sequences, and the final column shows the corresponding Weber wave.
    }
    \label{fig:Fig5-BWG}
\end{figure}

Our spectral contraction acts on the separated differential equations and their separation spectra.
The separated differential equations determine their solutions up to global multiplicative constants, so we understand the modal limits below up to overall normalization.
Figures~\ref{fig:Fig2-HG}--\ref{fig:Fig5-BWG} show the four spectral contractions through normalized real field profiles at $z=0$.
The top and bottom rows show the two parity sectors selected for each contraction, the first three columns show Gaussian modes along each contraction sequence, and the final column shows the corresponding transverse Helmholtz wave.
The Gaussian modes are square integrable, while the ideal fixed-shell Helmholtz waves have generalized Dirac-$\delta$ normalization and infinite transverse extent.
We normalize all profiles to unit maximum absolute amplitude on the fixed window $Q_{x},Q_{y}\in[-30,30]$.
We compare the last Gaussian mode in each row with its Helmholtz-wave counterpart using the Pearson correlation $\lvert C\rvert\in[0,1]$ and the normalized Frobenius error $E_{\mathrm{F}}\geq0$~\cite{DominguezCruz2026}.
Values $\lvert C\rvert\simeq1$ identify the same spatial pattern up to normalization, while $E_{\mathrm{F}}\simeq0$ identifies a small pointwise difference between the normalized profiles.
The Pearson correlation and the normalized Frobenius error compare spatial patterns on the normalization window without making a comparison of global amplitude, optical power, or total transverse energy.

The real Hermite-Gauss modes separate in Cartesian coordinates,
\begin{align}
    \Psi_{\boldsymbol{\sigma}, n_{x}, n_{y}}^{(\mathrm{HG})}(q_{x},q_{y}) =&~ h_{\sigma_{x}, n_{x}}(q_{x}) h_{\sigma_{y}, n_{y}}(q_{y}),
\end{align}
where $h_{\sigma_{a},n_{a}}(q_{a})$ is the real one-dimensional Hermite-Gaussian factor with parity $\sigma_{a}$ and modal number $n_{a}$.
The two factors obey the one-dimensional Hermite equations,
\begin{align}\label{eq:1D_HG}
    \begin{aligned}
        \left[ \hat{p}_{x}^{2} + \hat{q}_{x}^{2} - ( 2 n_{x} + 1 ) \right] h_{\sigma_{x}, n_{x}}(q_{x}) =&~ 0, \\
        \left[ \hat{p}_{y}^{2} + \hat{q}_{y}^{2} - ( 2 n_{y} + 1 ) \right] h_{\sigma_{y}, n_{y}}(q_{y}) =&~ 0.
    \end{aligned}
\end{align}
After multiplication by $\kappa$, these equations contract under $\kappa \to 0^{+}$, $2 \kappa n_{x} \to \cos^{2} \vartheta$, and $2 \kappa n_{y} \to \sin^{2}\vartheta$, which gives $2 \kappa N \to 1$ through $N = n_{x} + n_{y}$, to the Cartesian plane-wave equations
\begin{align}\label{eq:Stand_waves}
    \begin{aligned}
        \left[ \hat{P}_{x}^{2} - \cos^{2}\vartheta \right] \mathcal{T}_{\sigma_{x}}(Q_{x}\cos\vartheta) =&~ 0, \\
        \left[ \hat{P}_{y}^{2} - \sin^{2}\vartheta \right] \mathcal{T}_{\sigma_{y}}(Q_{y}\sin\vartheta) =&~ 0,
    \end{aligned}
\end{align}
where
\begin{align}
    \mathcal{T}_{\mathrm{e}}(s) =&~ \cos s, \qquad
    \mathcal{T}_{\mathrm{o}}(s) = \sin s.
\end{align}
In the limit, we obtain the parity-adapted Cartesian wave
\begin{align}
    \Phi_{\boldsymbol{\sigma},\vartheta}^{(\mathrm{C})}(Q_{x},Q_{y}) =&~ \mathcal{T}_{\sigma_{x}}(Q_{x}\cos\vartheta) \mathcal{T}_{\sigma_{y}}(Q_{y}\sin\vartheta).
\end{align}

The large-order Hermite asymptotics give the local convergence of each factor to its parity-selected trigonometric counterpart~\cite{NISTBOOK}.
For $0 < \vartheta < \pi / 2$, both Cartesian modal numbers diverge while their scaled values determine the single Cartesian separation angle $\vartheta$ and preserve the Cartesian parity vector $\boldsymbol{\sigma}$.
At $\vartheta = 0$ and $\vartheta = \pi / 2$, we retain only the nonzero parity-compatible factors.
Diagonal Hermite-Gauss modes contract to the same equations in Cartesian axes rotated by $45^{\circ}$.

Figure~\ref{fig:Fig2-HG} shows the Hermite-Gauss contraction for $\vartheta=\pi/4$.
The even-even and even-odd rows target the real parity-adapted Cartesian waves,
\begin{align}
    \begin{aligned}
        \Phi_{(\mathrm{e},\mathrm{e}),\pi/4}^{(\mathrm{C})} =&~ \cos \left( \frac{Q_{x}}{\sqrt{2}} \right) \cos \left( \frac{Q_{y}}{\sqrt{2}} \right), \\
        \Phi_{(\mathrm{e},\mathrm{o}),\pi/4}^{(\mathrm{C})} =&~ \cos \left( \frac{Q_{x}}{\sqrt{2}} \right) \sin \left( \frac{Q_{y}}{\sqrt{2}} \right),
    \end{aligned}
\end{align}
respectively.
The even sequence uses $n_{x}=n_{y}=N/2$ with $N\in\{8,16,104\}$, while the odd sequence uses $n_x=(N-1)/2$ and $n_y=(N+1)/2$ with $N\in\{9,17,105\}$.
Both sequences satisfy $2\kappa(n_{x}+n_{y})=1$, while the two scaled modal numbers approach $1/2$ with their Cartesian parities preserved.
The last even and odd Hermite-Gauss modes give $(\lvert C\rvert,E_{\mathrm{F}})=(0.9994,0.0367)$ and $(0.9958,0.0913)$, respectively.

The real Laguerre-Gauss modes separate in polar coordinates, $x+iy=\rho e^{i\theta}$,
\begin{align}
    \Psi_{\sigma, p, \ell}^{(\mathrm{LG})}(q_{\rho},q_{\theta}) =&~ l_{p,\ell}(q_{\rho})\mathcal{T}_{\sigma}(\ell q_{\theta}),
\end{align}
where $l_{p,\ell}(q_{\rho})$ is the real radial Laguerre-Gaussian factor with the radial modal number $p$, and $\ell$ is the azimuthal modal number.
The radial and angular factors obey
\begin{align}
    \begin{aligned}
        \left[ \hat{p}_{q_{\rho}}^{2} - i \hat{q}_{\rho}^{-1} \hat{p}_{q_{\rho}} + \ell^{2}\hat{q}_{\rho}^{-2} - 4p - 2 \ell - 2 + \hat{q}_{\rho}^{2} \right] l_{p,\ell}(q_{\rho}) =&~ 0, \\
        \left[ 4 \hat{J}_{y}^{2} - \ell^{2} \right] \mathcal{T}_{\sigma}(\ell q_{\theta}) =&~ 0.
    \end{aligned}
\end{align}
After multiplication by $\kappa$, the radial equation contracts under $\kappa \to 0^{+}$ and $2 \kappa( 2 p + \ell) \to 1$ at fixed $(\sigma, \ell)$ to the radial Bessel equation, while the angular equation remains unchanged through $q_{\theta} = Q_{\theta}$ and $4 \hat{J}_{y}^{2} = \hat{L}_{z}^{2}$,
\begin{align}
    \begin{aligned}
        \left[
            \hat{P}_{Q_{\rho}}^{2}
            - i\hat{Q}_{\rho}^{-1}\hat{P}_{Q_{\rho}}
            + \ell^{2}\hat{Q}_{\rho}^{-2}
            - 1
        \right] J_{\ell}(Q_{\rho}) =&~ 0, \\
        \left[\hat{L}_{z}^{2}-\ell^{2}\right]
        \mathcal{T}_{\sigma}(\ell Q_{\theta}) =&~ 0.
    \end{aligned}
\end{align}
The resulting real Bessel wave is
\begin{align}
    \Phi_{\sigma,\ell}^{(\mathrm{B})}(Q_{\rho},Q_{\theta}) =&~ J_{\ell}(Q_{\rho}) \mathcal{T}_{\sigma}(\ell Q_{\theta}),
\end{align}
where $J_{\ell}(Q_{\rho})$ is the Bessel function of the first kind with the azimuthal order $\ell$.

The Mehler-Heine asymptotics give the local convergence of the radial Laguerre-Gaussian factor to $J_{\ell}(Q_{\rho})$~\cite{NISTBOOK}.
The radial modal number $p$ diverges to fix the transverse-wavenumber shell, while the second modal number $\ell$ becomes the Bessel separation parameter and the parity $\sigma$ is preserved.

Figure~\ref{fig:Fig3-LG} shows the Laguerre-Gauss contraction for the fixed azimuthal modal number $\ell=6$.
The even and odd rows target the real Bessel waves,
\begin{align}
    \Phi_{\sigma,6}^{(\mathrm{B})}(Q_{\rho},Q_{\theta}) =&~ J_{6}(Q_{\rho})\mathcal{T}_{\sigma}(6Q_{\theta}), \qquad
    \sigma \in \{ \mathrm{e}, \mathrm{o}\},
\end{align}
using $N \in \{8, 16, 104\}$ and $p = ( N - 6 ) / 2$ in both parity sectors.
The radial modal number $p$ increases to fix the transverse-wavenumber shell, while $\ell=6$ becomes the Bessel separation parameter and the parity $\sigma$ is preserved.
The last even and odd Laguerre-Gauss modes give $(\lvert C\rvert,E_{\mathrm{F}})=(0.9991,0.0416)$ and $(0.9991,0.0421)$, respectively.

The real Ince-Gauss modes separate in elliptic coordinates, $x+iy=f\cosh(\xi+i\eta)$, 
\begin{align}
    \Psi_{\sigma, N, m}^{(\mathrm{IG})}(q_{\eta},q_{\xi}) =&~ Y_{\sigma,m}^{(N)}(q_{\eta}) M_{\sigma,m}^{(N)}(q_{\xi}),
\end{align}
where $Y_{\sigma,m}^{(N)}(q_{\eta})$ is the real angular Ince factor, and $M_{\sigma,m}^{(N)}(q_{\xi})$ is its real modified Ince counterpart.
The factors carry the parity $\sigma$, the total modal order $N$, the elliptic modal number $m$, and the dimensionless half-interfocal distance $q_{f}$.
They obey the angular and modified Ince equations,
\begin{align}
    \begin{aligned}
        \left[ \hat{p}_{q_{\eta}}^{2} - i q_{f}^{2}\sin(2\hat{q}_{\eta})\hat{p}_{q_{\eta}} - \alpha_{\sigma,N,m}(q_{f}^{2}) + N q_{f}^{2}\cos(2\hat{q}_{\eta}) \right] Y_{\sigma,m}^{(N)}(q_{\eta}) =&~ 0, \\
        \left[ \hat{p}_{q_{\xi}}^{2} + i q_{f}^{2}\sinh(2\hat{q}_{\xi})\hat{p}_{q_{\xi}} + \alpha_{\sigma,N,m}(q_{f}^{2}) - N q_{f}^{2}\cosh(2\hat{q}_{\xi}) \right] M_{\sigma,m}^{(N)}(q_{\xi}) =&~ 0.
    \end{aligned}
\end{align}
Under $\kappa \to 0^{+}$, $2 \kappa N \to 1$, and $q_{f}^{2} = \kappa Q_{f}^{2}$ at fixed $(\sigma, m, Q_{f})$, these equations contract to the Mathieu equations
\begin{align}
    \begin{aligned}
        \left[ \hat{P}_{Q_{\eta}}^{2} - a_{\sigma,m}(Q_{f}^{2}/4) + \frac{1}{2} Q_{f}^{2} \cos(2\hat{Q}_{\eta}) \right] Y_{\sigma,m}(Q_{\eta}) =&~ 0, \\
        \left[ \hat{P}_{Q_{\xi}}^{2} + a_{\sigma,m}(Q_{f}^{2}/4) - \frac{1}{2} Q_{f}^{2} \cosh(2\hat{Q}_{\xi}) \right] M_{\sigma,m}(Q_{\xi}) =&~ 0,
    \end{aligned}
\end{align}
where the separation spectra contract according to
\begin{align}\label{eq:IG--M_eigen}
    \lim_{\substack{\kappa \to 0^{+} \\ N \to \infty}} \alpha_{\sigma,N,m}(\kappa Q_{f}^{2}) =&~
    a_{\sigma,m}\left( \frac{1}{4} Q_{f}^{2} \right).
\end{align}
We obtain the real Mathieu wave as
\begin{align}
    \Phi_{\sigma,m}^{(\mathrm{M})}(Q_{\eta},Q_{\xi}) =&~ Y_{\sigma,m}(Q_{\eta})M_{\sigma,m}(Q_{\xi}),
\end{align}
where $Y_{\sigma,m}(Q_{\eta})$ and $M_{\sigma,m}(Q_{\xi})$ are the angular and radial Mathieu functions, respectively. 

Along parity-compatible modal orders, $N$ diverges to fix the transverse-wavenumber shell, while the second modal number $m$ becomes the Mathieu separation parameter and the parity $\sigma$ is preserved.

Figure~\ref{fig:Fig4-IG} shows the Ince-Gauss contraction for the fixed elliptic modal number $m=6$ and the scaled half-interfocal distance $Q_{f}=8$.
The even and odd rows target the real Mathieu waves
\begin{align}
    \Phi_{\sigma,6}^{(\mathrm{M})}(Q_{\eta},Q_{\xi})=Y_{\sigma,6}(Q_{\eta})M_{\sigma,6}(Q_{\xi}), \qquad
    \sigma\in\{\mathrm{e},\mathrm{o}\},
\end{align}
using $N\in\{8,16,104\}$, $2\kappa N=1$, and $q_{f}^{2}=\kappa Q_{f}^{2}$ in both parity sectors.
The even Ince separation constants take the values $43.6323$, $41.7766$, and $40.7824$, approaching the Mathieu value $a_{\mathrm{e},6}(16)=40.6609$, while the odd values $42.1668$, $40.1882$, and $39.3790$ approach $a_{\mathrm{o},6}(16)=39.3150$.
The last even and odd Ince-Gauss modes give $(\lvert C\rvert,E_{\mathrm{F}})=(0.9990,0.0456)$ and $(0.9991,0.0430)$, respectively.

The real Boyer-Wolf-Gauss modes separate in parabolic coordinates~\cite{Boyer1975p2215,Tschernig2024p5301}, $x+iy=(u+iv)^{2}/2$,
\begin{align}
    \Psi_{\sigma, n, r}^{(\mathrm{BWG})}(q_{u},q_{v}) =&~ U_{\sigma,n,r}(q_{u})V_{\sigma,n,r}(q_{v}),
\end{align}
where $U_{\sigma,n,r}(q_{u})$ and $V_{\sigma,n,r}(q_{v})$ are the real sextic-oscillator factors, $n$ is the astigmatic modal order, and $r$ is the Boyer-Wolf branch index.
The factors obey
\begin{align}
    \begin{aligned}
        \left[ \hat{p}_{q_{u}}^{2} - ( 2n + 3 )\hat{q}_{u}^{2} + \hat{q}_{u}^{6} - 2\beta_{n,r}
        \right]U_{\sigma,n,r}(q_{u}) =&~ 0, \\
        \left[ \hat{p}_{q_{v}}^{2} - ( 2n + 3 )\hat{q}_{v}^{2} + \hat{q}_{v}^{6} + 2\beta_{n,r} \right] V_{\sigma,n,r}(q_{v}) =&~ 0.
    \end{aligned}
\end{align}
After multiplication by $\sqrt{\kappa}$, these equations contract under $\kappa \to 0^{+}$, $2 \kappa n \to 1$, and $q_{c} = \kappa^{1/4} Q_{c}$ for $c\in\{u,v\}$ to the Weber equations
\begin{align}
    \begin{aligned}
        \left[\hat{P}_{Q_{u}}^{2}-\hat{Q}_{u}^{2}+2b\right] U_{\sigma,b}(Q_{u}) =&~ 0, \\
        \left[\hat{P}_{Q_{v}}^{2}-\hat{Q}_{v}^{2}-2b\right] V_{\sigma,b}(Q_{v}) =&~ 0,
    \end{aligned}
\end{align}
with separation-spectrum contraction along branch sequences,
\begin{align}
    b =&~ \lim_{\substack{\kappa \to 0^{+} \\ n \to \infty}} \left[-\sqrt{\kappa} \, \beta_{n,r_{n}} \right],     \qquad
    r_{n}\in\mathcal{R}_{n}.
\end{align}
The contracted factors give the real Weber wave
\begin{align}
    \Phi_{\sigma,b}^{(\mathrm{W})}(Q_{u},Q_{v}) =&~ U_{\sigma,b}(Q_{u}) V_{\sigma,b}(Q_{v}),
\end{align}
where $Q_{x} + i Q_{y} = (Q_{u} + i Q_{v})^{2}/2$, $U_{\sigma,b}(Q_{u})$ and $V_{\sigma,b}(Q_{v})$ are the real Weber factors, and $b\in\mathbb{R}$ is their continuous parabolic separation parameter.

The diverging modal order $n$ fixes the transverse-wavenumber shell, while the branch sequence $r_{n}$ selects the continuous Weber separation parameter $b$ and the parity-compatible modal orders preserve $\sigma$.

Figure~\ref{fig:Fig5-BWG} shows the Boyer-Wolf-Gauss contraction for the fixed Weber separation parameter $b=2$.
The even and odd rows target the real Weber waves,
\begin{align}
    \Phi_{\sigma,2}^{(\mathrm{W})}(Q_{u},Q_{v})=U_{\sigma,2}(Q_{u}) V_{\sigma,2}(Q_{v}), \qquad
    \sigma\in\{\mathrm{e},\mathrm{o}\}.
\end{align}
The even sequence uses $(n,r_{n})=(8,-1),(16,-2),(162,-5/2)$, for which $-\sqrt{\kappa_{n}} \beta_{n,r_{n}}=1.1514,2.4130,2.0060$, while the odd sequence uses $(n,r_{n})=(11,-3/2),(15,-3/2),(359,-5/2)$, for which $-\sqrt{\kappa_{n}}\beta_{n,r_{n}}=2.2361,2.0569,1.9994$.
In both sectors, $2\kappa_{n}n=1$, the modal order fixes the transverse-wavenumber shell, and the parity-compatible branch sequence selects the limiting value $b=2$.
The last even and odd Boyer-Wolf-Gauss modes give $(\lvert C\rvert,E_{\mathrm{F}})=(0.9991,0.0417)$ and $(0.9996,0.0274)$, respectively.

Our spectral contraction connects the Gaussian-beam families with their Helmholtz-wave counterparts at the separated differential equation and separation-spectrum levels.

\section{Conclusion}
\label{sec:Sec5}
We introduced the optical scale ratio as the physical parameter governing the contraction from separable Gaussian beams to transverse Helmholtz waves.
Its square root compares the transverse Helmholtz length with the Gaussian transverse scale.
For a transverse Helmholtz shell chosen within the paraxial angular spectrum, the large-Rayleigh-range limit expands the Gaussian beam, suppresses its transverse rescaling and wavefront curvature, and drives our optical scale ratio towards zero.
Coupling this change of scale to the higher-modal-order limit preserves the shell and transforms the isotropic and astigmatic oscillator eigenvalue problems of the paraxial regime into the transverse Helmholtz unit-shell problem.

As our optical scale ratio vanishes, it realizes the In\"{o}n\"{u}-Wigner contraction from the isotropic Gaussian-beam $\mathrm{su}(2)$ algebra to the transverse Helmholtz $\mathrm{e}(2)$ algebra; conversely, its finite nonzero values parametrize the Cayley-Klein deformation.

The joint limit of the vanishing scale ratio and the higher modal order produces a spectral contraction in which the two modal numbers of each Gaussian beam fix the transverse-wavenumber shell and determine the single separation parameter of its corresponding transverse Helmholtz wave, while preserving parity.
We establish the Hermite-Gaussian to parity-adapted Cartesian, Laguerre-Gaussian to Bessel, Ince-Gaussian to Mathieu, and Boyer-Wolf-Gaussian to Weber contractions at the wave-operator, separation-operator, separated-differential-equation, and separation-spectrum levels.

Our optical scale ratio gives physical and modal content to the algebraic contraction and extends the Cartesian and polar connection to the four separable coordinate families in the paraxial and Helmholtz regimes.



\section*{Acknowledgment}
B.~M.~R.~L. thanks Jacinta Alderete Galan for providing daycare support throughout this work.
He also acknowledges support and hospitality as an affiliate visiting colleague at the Department of Physics and Astronomy, University of New Mexico.

\section*{Data Availability Statement}
All data are available from the corresponding author upon reasonable request.



%

\end{document}